# Understanding Electronic Peculiarities in Tetragonal FeSe as Local Structural Symmetry Breaking

Zhi Wang[1], Xin-Gang Zhao[1], Robert Koch[2], Simon J. L. Billinge[2,3] and Alex Zunger[1*].

[1]University of Colorado, Renewable & Sustainable Energy Institute, Boulder, Colorado 80309, USA.

[2]Condensed Matter Physics and Materials Science Department, Brookhaven National Laboratory, Upton, NY 11973.

[3]Department of Applied Physics and Mathematics, Columbia University, New York, New York 10027, USA.

*Corresponding author: **Alex Zunger.** alex.zunger@colorado.edu

## Abstract

Traditional band theory of perfect crystalline solids often uses as input the structure deduced from diffraction experiments; when modeled by the minimal unit cell this often produces a spatially averaged model. The present study illustrates that this is not always a safe practice unless one examines if the intrinsic bonding mechanism is capable of benefiting from the formation of a distribution of lower symmetry local environments that differ from the macroscopically averaged structure. This can happen either due to positional, or due to magnetic symmetry breaking. By removing the constraint of a small crystallographic cell, the energy minimization in the density functional theory finds atomic and spin symmetry breaking, not evident in conventional diffraction experiments but being found by local probes such as atomic pair distribution function analysis. Here we report that large atomic and electronic anomalies in bulk tetragonal FeSe emerge from the existence of distributions of local positional and magnetic moment motifs. The found symmetry-broken motifs obtained by minimization of the internal energy represent what chemical bonding in tetragonal phase prefers as an intrinsic energy lowering (stabilizing) static distortions. This explains observations of band renormalization, predicts orbital order and enhanced nematicity, and provides unprecedented close agreement with spectral function measured by photoemission and local atomic environment revealed by pair distribution function. While the symmetry-restricted strong correlation approach has been argued previously to be the exclusive theory needed for describing the main peculiarities of FeSe, we show here that the symmetry-broken mean-field approach addresses numerous aspects of the problem, provides intuitive insight into the electronic structure, and opens the door for large-scale mean-field calculations for similar *d*-electron quantum materials.

## I. Introduction

Traditional electronic band theory of crystalline solids has subscribed to the Bloch [1] and Bragg [2] paradigms, whereby the minimum-size crystallographic unit cell obtained from standard refinement [3] of X-ray Bragg diffraction (XRD) [4] data provides an all-encompassing description of the atomic arrangement used as input in band theory. However, it is becoming apparent that some defect-free, stoichiometric crystalline materials show local, static motifs, which on a local scale are inconsistent with the observed XRD crystal symmetry [5–10] because such local features are, by their nature, not periodic. A classic example of static microscopic local environments is the atomic-scale compositional configurations common in random $A_xB_{1-x}$ *alloys* of composition x, established by the numerous ways to locally configure an A-site by different number of A and B neighbors, and *vice versa*, leading effectively to large unit cell [9–12]. But could such polymorphous network (*i.e.*, showing a distribution of different local environments) *exist as intrinsic description of ordered, non-alloyed crystal?* This might be expected in solids where the units forming the local chemical bonds or local magnetic configurations have the adaptive ability to be stable in numerous local configurations. Examples of crystalline solids implicated in forming a distribution of local *atomic displacements* ("*positionally polymorphous*") include bond disproportionation in crystalline nickelates [13], $BaBiO_3$ [14] and $CsAuCl_3$ [15], or the effect of octahedral rotations [16] on the electronic structure of nominally cubic halide perovskites [17]. Examples of crystalline solids forming local *magnetic moments* ("*spin polymorphous*", where local magnetic moments persist in the paramagnetic phase) include Mott oxides [18–20] and cuprate superconductors [21,22].

The significance of the fact that such local symmetries cannot be meaningfully reduced to a trivial minimal average cell, lies in the fact that the electronic structure can be sensitive to local distortions even in periodic solids. For example, the recent minimal-cell mean-field DFT by Long *et al.* [23] could not explain the observed anomalies in tetragonal FeSe such as band narrowing and local atomic nematicity. This is the case when the minimal-cell model represents a macroscopically averaged, high-symmetry approximation $S_0 = \langle S_i \rangle$ of all local, low-symmetric environments $\{S_i;\ i = 1, N\}$. Thus, the calculated property $\langle P \rangle = P(S_0)$ of the averaged configuration $S_0$ (as commonly done in virtual crystal and coherent potential approximations in alloy theory) could deviate [24–26] from the correct $P_{\text{obs}} = \Sigma P(S_i)$ being the average of the properties of all microscopic configurations. When such discrepancies between the prediction of minimal (usually a single or a few formula units per cell, f.u./cell) band theory and experiment were noted -- such as the celebrated absence of band gaps in oxide Mott insulators -- dynamic electron-electron correlation in a symmetry constrained structure were often introduced as the essential cure [27,28].

Indeed, standard theoretical approaches (*e.g.*, density functional theory, DFT) based on the Bragg-Bloch minimal crystallographic unit cell [29–33] illustrated in Fig. 1a showed a number of striking peculiarities in tetragonal FeSe relative to experiment: (i) Significant (by a factor of 3) narrowing of the bands near the Fermi level has been observed in angle-resolved photoemission spectroscopy (ARPES) of the tetragonal phase of FeSe [34–36] (illustrated by the empty red circles and vertical red arrow in Fig. 1a). Such band renormalization relative to the standard minimal-cell reference band structure of tetragonal FeSe cell (blue lines and vertical blue arrow in Fig. 1a), has been argued to evidence the exclusive role of strong electronic correlation. (ii) The experimentally observed electronic structure has lower apparent symmetry ('nematicity') [34–39], absent from simple band theory of the minimal-cell model [29–31]. It has been argued [40], on the basis that the observed *average* structural symmetry breaking is too small in the macroscopically averaged minimal cell to account for the observed symmetry removal, that this nematicity must be a result of *electronic* symmetry breaking [29–31], and is exclusively driven by strong electronic correlations. However, this does not suffice for FeSe [40], where a key complexity emerges: The coexistence of *electronic* correlation in symmetry-restricted structure (minimal-cell model), concomitantly with *structural symmetry breaking* that is not explainable by the paradigm of symmetry conservation. Whereas the *average* structural symmetry breaking is small, *local* structural probes [41–43] reveal static and significant local orthorhombicity that remains theoretically unexplained by the symmetry-restricted models. Such local structure probes offer the opportunity to go beyond the minimal unit cell picture, allowing researchers to investigate how the local atomic structure, over a few Angstroms to nanometers, deviates from the average XRD structure. Successful theories of atomic and electronic structure must then be examined by their agreement with such local probes.

The reference electronic structure of Fig. 1a was calculated in the standard tetragonal minimal-cell model, and hence it cannot break symmetry by creating a distribution of positional and/or spin local environments, even with improved DFT exchange-correlation functionals, on account of the highly symmetric geometry imposed on it by the Bragg-Bloch band paradigm. Rather than leapfrog from mean-field theory to the framework of dynamic electronic correlations to interpret the peculiarities of FeSe, we opt to first examine the performance of mean-field DFT, un-obscured by approximations that do not represent what DFT can indeed do. Specifically, we wish to establish what are the minimal theory ingredients needed to explain the base properties of the paramagnetic tetragonal FeSe phase, allowing all types of local environment to find the lowest total internal energy, within a given macroscopic cell symmetry (here, tetragonal).

We find that considering only positionally polymorphous structures [17] or only spin polymorphous structures [18–20], is not adequate for FeSe, and that a complex *interrelation* between the two is needed (Table 1 later). We report in the tetragonal phase of bulk FeSe the unexpected findings that the *structurally induced* electronic symmetry removal with its attendant total energy lowering (Fig. 1e) leads to significant band narrowing (renormalization factor of 3 from blue to red vertical arrows in Fig. 1b), and nematic orbital ordering (colored contours in Fig. 1d). These are predicted by DFT calculations without imposing strong dynamic correlations, but only when enlarged unit cells (Fig. 1e) are used, large enough to encompass positional as well as spin symmetry breaking polymorphous networks. The resulting polymorphous network is tested by its ability to lower the total energy per atom as the unit cell is systematically enlarged (*i.e.*, a static, non-thermal effect); converging after a minimum size, encompassing the variational distribution of bonds is reached (Fig. 1e). This results in a distribution of local moments in the paramagnetic phase, as well as a distribution of Fe-Se and Se-Se bond lengths, and different charges on different Fe atoms. This picture of combined positional and spin local environments is then tested by computing from first principles the (1) Pair Distribution Function (PDF), and (2) the observed electronic structure, finding good agreement with room-temperature experiments without fitting structural parameters or imposing strong correlations: The polymorphous network, having a distribution of local structural and spin motifs, reproduces the experimental PDF (Fig. 2c) and experimental ARPES (comparison between the colored contour and the red empty circles in Fig. 1b) in tetragonal FeSe, all being the direct results of the polymorphous nature of the structure. DFT calculations using instead the *averaged, monomorphous minimal-cell model* to predict electronic properties (Fig. 1a,c) miss all these pictures. However, such model failed not necessarily because of the absence of strong electron-electron correlation, but because unlike the averaged, minimal-cell view, the tetragonal phase of FeSe consists of distributions of different local symmetry-broken motifs. We note that the static energy-lowering positional polymorphous network cannot be easily depicted by the standard electron-phonon interaction, because it is highly anharmonic and not easily expressible as a sum of a few phonon states. It is treated in this work by performing a static minimization of all forces on atoms in the supercell (as described in the next section), thus being inherently anharmonic and multi-phononic.

The broad implication of this study is that even in the pure, ordered, stoichiometric compounds, simply picking up a crystal structure from structural or magnetic databases then proceeding with calculations of the band structure in the traditional Bragg-Bloch paradigm is not always a safe practice. This study provides insight into the electronic structure by considering the effect of structural and spin degrees of freedom on bonding and hybridization. The symmetry-broken mean-field band theory addresses numerous observed peculiarities, which have been exclusively attributed to symmetry-restricted strong correlation effects. Beyond the theoretical importance of discerning the minimal mechanism at play -- symmetry breaking vs. dynamic correlation -- the practical significance of this realization is that it opens the door for large-scale calculations of properties of similar *d*-electron quantum materials important in catalysis, thermoelectrics and photovoltaics that could be prohibitively complex with correlated realistic models where polymorphous networks are likely to abound.

## II. Results

***Allowing a paramagnetic phase to lower its energy via polymorphous networks of spins and displacements:*** The paramagnetic (PM) phase of tetragonal FeSe is modeled by a supercell that has global tetragonal lattice symmetry and (as appropriate for a paramagnet) a zero net magnetization, but the local magnetic moment as well as local atomic displacements can develop nonzero values if these minimize the total mean-field (DFT) energy. Indeed, the difference between antiferromagnetic (AFM) and PM orders is not that the later phase has vanishing moments on an atom by atom basis, but rather that the PM has short range (local) order (SRO) rather than long range order (LRO) as in AFM. In dynamic mean-field theory, paramagnetism is described [28] by allowing a single ion interacting with a spatially average mean-field 'bath' to sustain time dependent spin fluctuations, thus averaging its moment over time to zero. Here we adopt instead a picture with a *spatial* distribution of up and down spins over a large supercell averaging *spatially* to zero net magnetization. Ergodicity implies that in principle a large enough cell will manifest all local environments visited by an alternative time domain method having but a single Fe atom fluctuating in time. This representation allows the formation of different spin local environments -- such as up-spin Fe coordinated locally by variable number of up- and down-spin Fe atoms (so called, spin polymorphous network). The representation also provides the possibility of creating broken spatial symmetries (positional local environments), if these would lower the total energy. These include distribution of Fe-Fe and Se-Se bond lengths, local deviations from tetragonal symmetry (so called, positional polymorphous network), as well as different moments and different charge density on different Fe sites. The cell size is increased to achieve convergence in the total energy per atom (Fig. 1e). If the underlying chemical bonding in the system at hand does not benefit from the existence of a distribution of different local environments, the total energy will not be lowered. This is the case for the isovalent ZnSe system having a unique minimal unit cell of a single formula unit, as the single motif of Zn coordinated by four equivalent Se sites suffices to lower the total energy.

In the present approach, the Fe sites are occupied by up-spin and down-spin (collinear magnetization) Fe atoms; spin occupation of the lattice sites is selected using the "Special Quasirandom Structure" (SQS) method [11] to best mimic the high-temperature limit of the spin-spin correlation functions of an infinite sized supercell. Such method, used extensively in the past to model chemical alloys $A_xB_{1-x}$ of composition x, is extended here to spins. Agreement with experiment can presumably be improved by including the short-range spin order in a finite-temperature PM phase, analogous to SRO in chemical alloys [44,45]. Details about the supercell generation have been given in Appendix A. The method is validated by internal convergence tests, as well as by comparison of the local positional distribution with accurate experimental PDF [41], and by comparing local distortions to models such as "local orthorhombicity" [41] $\eta(R)$ (which measures the distortion on the local tetragonal structure at each Fe site).

***Identification of two concomitant types of polymorphous networks by energy minimization***. Seeking energy minimum reveals two types of symmetry breaking degrees of freedom related to the electronic structure of tetragonal FeSe: the local magnetic moments $\{\mu_i\}$ on Fe sites, and the local atomic displacements $\{\Delta\boldsymbol{R}_i\}$ for Fe and Se atoms. The global structure emerges from the collection of these local motifs. Both sets of degrees of freedom are determined by DFT total energy minimization using tetragonal supercell large enough to accommodate symmetry breaking, should it lower the total energy. In a *spin polymorphous description*, the PM structure is allowed to have a distribution of non-zero local magnetic moments $\mu_i \neq 0$, adding up to a zero global magnetic moment $\sum_i \mu_i = 0$, defining the PM phase [18,19]. In a *positional polymorphous description*, atoms are allowed to develop local displacement $\Delta\boldsymbol{R}_i \neq 0$ subject to the constraint of a global tetragonal cell $\sum_i \Delta\boldsymbol{R}_i = 0$. Such dual spin and positional polymorphous descriptions require cells larger than minimal unit cells (*i.e.*, supercells). If, on the other hand, the assumed repeating unit cell is the nominal, smallest cell (for FeSe this is a primitive, 2-f.u. cell), then the PM state is described, by default, as a non-magnetic (NM) structure, where all local moments are taken as zero $\mu_i = 0$, and the positional degrees of freedom are restricted to the ideal, undisplaced values $\Delta\boldsymbol{R}_i = 0$.

Table 1 summarizes the results for different cells using the "yes" or "no" descriptions of the spin and positional polymorphism.

The (**No, No**) state (1) corresponds to the often-used monomorphous NM minimal-cell model -- a fully restricted DFT model using the minimal unit cell having zero magnetic moment and no atomic displacements on each and every site [29–33]. Table 1 shows that this picture leads to a very high total energy (292 meV/f.u. above the reference PM state), large errors in describing the experimentally observed PDF (Fig. 2a), no local orthorhombicity, no electronic nematicity (Fig. 1c), being in qualitative disagreement with the observed ARPES (Fig. 1a).

The (**No, Yes**) state (2) allows atomic displacements off the high-symmetric Wyckoff positions, lowering the total energy by a small amount (8 meV/f.u.), but in the absence of magnetic moment on all sites, the electronic subsystem is still nonmagnetic, unable to couple to the structural symmetry breaking, thus failing to show the nematicity, nor agreement with PDF or ARPES.

States (3) and (4) allow magnetic moment formation in the absence (**Yes, No**), and in the presence (**Yes, Yes**), respectively, of atomic displacement. Both states show large, constant lowering of the total energy (about 300 meV/f.u.) due to magnetic moment formation, which is consistent with Fig. 1e. However, only if local magnetism is accompanied by positional polymorphism -- as in the (Yes, Yes) state (4), which develops local atomic displacement -- can the electronic structure couple to the symmetry breaking, lowering further the total energy. State (4) is 37 meV/f.u. lower than state (3), and yields not only the best PDF (R-factor error = 6.5%, remarkable agreement for a fully first principles PDF), but also the electronic nematicity, and a band structure showing good agreement with ARPES (Fig. 1b,d and Fig. 2). Note that in Fig. 1e, the energy lowering from minimal cell (x-axis = 2) to large cells is initially due to spin polymorphism (magnetic relative to nonmagnetic), whereas subsequent energy lowering is the usual size convergence due to adjustment in atomic positions. We give in Appendix A details about the 4 states.

***Ab initio reproduction of PDF measurement of local atomic scale structure resulting from interaction of positional and spin polymorphous networks:*** Applying local structure probe to tetragonal FeSe such as PDF by Frandsen *et al.* [42], Konstantinova *et al.* [43] and by Koch *et al.* [41] showed that least-square fitting of the data in a phenomenological model required assuming low-symmetry structure such as local orthorhombic distortions in the tetragonal phase. This cannot be realized by a small unit cell periodic structure and thus such a virtual structure cannot be used as input to band theory to determine the ensuing electronic properties. The polymorphous supercell (Fig. 1b,d,e and Table 1) obtained from energy minimization, on the other hand, is an *a priori* real-space periodic structure, and can be used in band theory. We show in Fig. 2 the calculated PDF from (a) state (1) (**No, No**) in Table 1 and Fig. 1a, *i.e.*, the monomorphous nonmagnetic minimal-cell model (2 f.u./cell), (b) state (3) (**Yes, No**) in Table 1, *i.e.*, the constrained polymorphous PM supercell in the absent of atomic displacement (384 f.u./cell), and (c) state (4) (**Yes, Yes**) in Table 1 and Fig. 1b,d, *i.e.*, the relaxed polymorphous PM supercell in the present of atomic displacement (384 f.u./cell), all compared with room-temperature experimental PDF [41]. To minimize the periodicity error of the repeated, finite-size supercells, for PDF calculations we surround the *active* core 384 formula units of FeSe by additional, electronically frozen bulk-like tetragonal FeSe; such 'padded' structure contains in total 3,072 formula units (6,144 atoms). We find that:

(1) The calculated PDF of the (No, No) monomorphous NM model gives poor agreement with experiment (Fig. 2a; error $R_w$=0.62 in the 0-5 Å local range and $R_w$=1.19 in the 5-50 Å long range). (2) However, we find that by only inducing the spin polymorphism, DFT for the (Yes, No) constrained PM supercell already gives a significantly better PDF (0.11 and 0.10 errors in the local range and long range, respectively), as shown in Fig. 2b. This is because DFT finds better lattice constants in the (Yes, No) supercell than the ones in the (No, No) minimal cell (see Table 2). But in Fig. 2b there are still anomalous features not seen in experiment (*e.g.*, the misfit between fit and measured marked by the red circle at around 4 Å [41]). (3) Finally, the (Yes, Yes) relaxed polymorphous PM supercell shown in Fig. 2c clearly is the best agreement with experiment, which has $R_w$=0.065 in the local range (almost 50 % improvement if compared with Fig. 2b), and $R_w$=0.10 in the long-range region, the same agreement as Fig. 2b. Appendix B gives the details for PDF calculations.

***Emergence of a distribution of bond length, inter planar Fe-Se separations and a distribution of local orthorhombicity:*** FeSe has an Fe-Se-Fe layered structure which is

orthorhombic (*Cmme*) below 90 K [46] and paramagnetic down to 0.4 K [47], and transforms to a tetragonal structure (*P4/nmm*) above 90 K [48]. Fig. 3a,b describe the average tetragonal structure as deduced from conventional XRD. There are 3 unique Se-Se distances, which are in plane, intra-layer, and inter-layer Se-Se (red, blue, and green dash lines in Fig. 3a). Fig. 3c shows the bond length distribution in the (Yes, Yes) polymorphous PM supercell (384 f.u./cell) as a function of atom-atom separation. Although the average XRD structure has single, unique Fe-Fe and Fe-Se, and 3 Se-Se bond distances (vertical dashed lines), energy minimization gives a full distribution of such bond lengths, showing clear non-Gaussian distributions, leading to good agreement with experimental PDF, and indicating that this is not due to thermal effects, but rather reflects the natural a-thermal preference of chemical bonding in this system.

Fig. 3d shows the local distance along [001] direction between the center Fe plane and the Se plane ("selenium height" [31]), denoted as $z_{Se}$, from XRD [49] and from the (Yes, Yes) polymorphous PM supercell. It has been argued in previous literature that monomorphous NM minimal-cell (No, No) DFT without strong correlations would give wrong $z_{Se}$ ($z_{Se}$=1.36 Å in previous literature [31]). However, it can be seen that now the polymorphous PM supercell shows $z_{Se}$ having a distribution from 1.37 Å to 1.7 Å (black line in Fig. 3d), in good agreement with the average $z_{Se}$ from XRD 1.473 Å (red vertical line in Fig. 3d). The significantly larger $z_{Se}$ in the PM supercell shows the effect of non-zero magnetic moment at local sites, while the large distribution of $z_{Se}$ indicates the large difference on potential of different local motifs.

The atomic positions obtained from DFT energy minimization of the (Yes, Yes) polymorphous PM supercell can also be analyzed to reveal the predominant local structural motifs. We find that the radius-dependent local orthorhombicity $\eta(R) = 2|a-b|/(a+b)$ [41] captures well the structural motifs. As shown in Fig. 3e, for large radius R>30 Å, $\eta(R)$=0, which denotes the macroscopic tetragonal phase, but looking at R in the 2-30 Å range reveals that the tetragonal structure can lower its energy by locally adopting a spread of orthorhombic distortions. Fig. 3e compares the calculated $\eta_{calc}(R)$ from the (Yes, Yes) PM supercell in Table 1 in a sphere $R_c$<R (see schematic plot in Fig. 3e insert) to the measured $\eta_{exp}(R)$ [41]. The agreement is excellent. Notably, both experimental and theoretical measures of the local distortions $\eta(R)$~2% are significantly larger than the values obtained from the average (XRD) distortions of 0% for 90 K and 0.12% for 84 K [41,46]. The latter small average distortions led to the argument [40] that structural symmetry breaking is too small to account for the observed electronic symmetry removal. Since this is not the case when one considers the pertinent local distortions (to which band structures respond), we will next examine the electronic response to the local distortions based on the supercell calculation.

***Distribution of magnetic moments and local charge densities:*** Fig. 4a,b show via red lines the distribution of local magnetic moments (a) and charges (b) on atoms in the (Yes, Yes) PM supercell, where the vertical blue lines denoted "NM prim cell" show the corresponding quantities in the fictitious (No, No) monomorphous NM minimal-cell model, being obviously very different. Whereas the net magnetic moment in the PM phase is zero, there is a nontrivial distribution of local moments of both orientations. Similarly, the distribution of atomic charges (computed by the projection of the occupied wavefunctions onto the spherical harmonics within a spherical region centered at Fe ions) shown in Fig. 4b clearly reveals the ability of FeSe to sustain a range of chemical bonding pattern, which is obviously absent in the isovalent compounds such as ZnSe.

***Consequence of local motifs for electronic properties***: ***symmetry removal and band narrowing.*** The $E$ versus $k$ band structure of a supercell does not lend itself to intuitive analysis as it appears as meaningless spaghetti. At the same time, the supercell representation is needed to afford symmetry breaking, unrestricted by geometrical constrain of a minimal cell. We therefore rigorously unfolded it [50–52] to the primitive BZ, providing an "Effective Band Structure" (EBS) -- a three-dimensional picture of the distribution of spectral density, including both coherent and incoherent contributions, all obtained from nominally mean field DFT [51,52] (basic concept of EBS is given in Appendix C).

Fig. 1 compares the measured ARPES [34–36] bands with the (No, No) monomorphous NM band structure results in (a) and the (Yes, Yes) polymorphous PM results in (b). As shown in Fig. 1a, the monomorphous NM model fails to reproduce the band structure: Neither the energies of $\alpha$-$\gamma$ states at Γ, nor the bandwidths of those states nearby Γ agree with experimental results. Such failure of the monomorphous model is often attributed to some fundamental failure of the DFT picture. However, the monomorphous approximation Fig. 1a is not the best that DFT can do. Fig. 1b shows that if we allow electrons to interact with the spin and positional degrees of freedom, without the restriction to high-symmetry small unit cell, we find a much richer picture and achieve a good agreement with ARPES results. Here, the spectral functions are calculated without ARPES matrix element effect, and obtained from, as we have, *random* short-range order in the PM phase (corresponding to very high temperature), which tend to overemphasize broad features, but we still can locate the three bands α, β and γ from a Lorentzian function analysis of the calculated spectral function along the M-Γ-M direction (see Appendix D). The calculated EBS from the polymorphous network not only reproduces the correct state energy and degeneracy splitting among $\alpha$-$\gamma$ at Γ, but also provides the correct bandwidths for each of the three bands. The band narrowing (renormalization) has been traditionally attributed to strong electron correlations on the basis of comparing with monomorphous NM minimal-cell DFT [29–31,34–36]. Here it is naturally explained by symmetry breaking and the ensuing inter band coupling sanctioned by DFT mean field. The spectral function shows fuzzy bands below 75 meV binding energy, while ARPES sees clear band dispersions there [34] after filtering-out the low-intensity signals and/or using second-derivative imaging. This may be due to the fact that the spin distribution function in the PM phase in this work is lacking possible spin short-range order [44,45].

In Fig. 1c,d we show the orbital order by plotting in *real space* the cross-sections (here, (001) plane at z=0) of the partial charge density in an energy region -0.03 eV < E-$E_F$ < 0.02 eV. The (No, No) monomorphous NM minimal-cell model results in an identical $C_4$-symmetric distribution around every and each Fe sites, indicating an equal orbital occupation $d_{xz}$:$d_{yz}$ = 1:1 of partial charge-density for all Fe atom (Fig. 1c). On the other hand, the (Yes, Yes) polymorphous PM structure, as shown in Fig. 1d, gives nematic partial charge-density around Fe atoms: Electron are localized in an orbital pointing mainly along x or y direction but not equivalently at Fe sites. Note that the orbital orders are drawn from bands within specific energy range and therefore do not necessarily present the overall crystal symmetry. Such local nematicity from single-determinant DFT without strong correlations is attributed to the existence of many local, low-symmetric motifs, which cannot be captured in a monomorphous minimal cell.

### III.     Discussion

The observed atomic and electronic structure of tetragonal FeSe has lower apparent symmetry than that of the macroscopic crystallographic structure. Here, for bulk tetragonal FeSe we carry out first-principles calculations on large supercells that preserve the observed global symmetry but do not artificially impose such local symmetry either on spin order or on atomic displacements. The significant finding of this study is that large anomalies in the atomic and electronic structures of the tetragonal phase FeSe can emerge from the existence of an intrinsic distribution of static, non-thermal, local positional as well as spin symmetry-broken motifs, mandated by the intrinsic chemical bonding as captured by DFT. Noting them in a calculation requires abandoning the traditional minimal unit cell picture. This symmetry-broken approach explored within mean-field DFT reveals rich phenomena including a nontrivial PDF, local nematicity, wavefunction symmetry removal, and concomitant band narrowing indicating mass renormalization, all in substantial agreement with experimental results at room temperature, but without imposing strong correlations. It shows that DFT calculations using traditional highly symmetric minimal-unit cells, where the symmetry breaking is in effect averaged over, pose an unnecessary restriction on mean-field theory. The local symmetry lowering observed experimentally, and now in DFT, is not just an approach, nor just a model for materials calculation, but represents the true nature of this class of materials that we are calling *polymorphous networks.* We conclude that the impression that the failure of mean-field DFT in previous studies [28–33,40], which has been argued to imply the necessity of strong dynamic correlation, may have been premature. The broader implication is that band structures cannot be automatically performed with the traditional, minimally sized XRD unit cell without examining if the local symmetry differs from the global average. This realization holds the potential of crossing the divides of individual sub discipline and could affect other areas of materials physics and chemistry where polymorphous networks are likely to abound.

## Appendix

### Appendix A: Computational methods.

**DFT method:** All calculations are done using spin-polarized density functional theory in a pseudopotential representation with augmented plane wave basis, using the exchange correlation functional PBE GGA+U and spin orbit coupling, all implemented by the VASP code. The plane wave basis cutoff energy is 320 eV. We select U=0.875 eV on Fe *d*-orbitals within the accepted range of values used previously. Van der Waals interaction has been involved in all calculations using opt86b method. Lattice parameters and total energy of each model have been shown in Table 2. To make the total energies between different cells comparable, we use an equivalent k-point mesh for total energy calculations in all cells. The paramagnetic supercell is generated following the special quasirandom structure (SQS) method using the ATAT code, the same procedure as applied in ref [18].

**Modeling of paramagnetic FeSe supercells.**

(i) The global shape of supercells is fixed to the macroscopically observed tetragonal symmetry.

(ii) The Fe sites in supercells are occupied by up- and down-spin (collinear) Fe atoms to achieve the closest simulation of a perfectly random (*i.e.* high-temperature limit) paramagnetic phase. The paramagnetic configurations are generated using Special Quasirandom Structure (SQS) method [11]. SQS method tries to reproduce pair and many-body correlation functions in the best way possible for a given supercell size $N$ [11]. The observable property $P$ calculated for such an SQS structure is not simply the property of a single snapshot configuration but approximates the ensemble average $P$ for the random configuration, as described in Refs. [11,53]. Furthermore, in general, SQS supercell with large size gives more reliable result than the ensemble average along many small random supercells, as shown by Ref. [53]. Convergence tests to total energy as a function of SQS size were tested as shown in main text Fig. 1e. Note that one can construct a paramagnetic state by using a time-dependent dynamic representation that is spatially homogeneous (*i.e.*, a single magnetic ion), or, as done here, using a time-independent static representation that is spatially inhomogeneous (supercell with many different local motifs). The system is a paramagnet in either cases—whether the time coordinate is computationally discretized (spin rotation in QMC) or if the position coordinate is computationally discretized (supercell).

(iii) The total energy minimization is performed by relaxing all internal atoms following ab initio forces while retaining the symmetry of the lattice vectors of the supercell (here, tetragonal). Atoms can be nudged initially to avoid trapping in local minima.

(iv) The paramagnetic configuration (which Fe site has up spin and which one has down spin) is fixed, *i.e.*, we do not allow spin flips.

(v) Wavefunctions are not symmetrized afterwards.

**Appendix B: PDF calculations**

**PDF from DFT.** All calculations for PDF from DFT-optimized structure are done using PDFgui software [54]. For tetragonal FeSe, $Q_{damp}$ and $Q_{broad}$ are fixed at 0.042 and 0.01, while $s_{ratio}$ and $r_{cut}$ are set to 1.0 and 0, respectively, the same values as used in ref [41]; scaling factor, $\delta_1$ and atomic displacement parameters (ADPs) are fitted by PDFgui (values listed in Table 3). For all models, all atomic positions and lattice constants are given by DFT total energy optimization without fitting or postprocessing. For every supercell, the PDF of short-range region (1.5-5 Angstrom) is calculated using exactly the DFT total-energy minimized atomic positions, while the PDF of long-range region (5-50 Angstroms) is calculated using the same parameters as the short-range PDF (atomic positions, scaling factor, $\delta_1$ and ADPs) but with a "padding" method, adding additional bulk-like tetragonal FeSe all around the central cell, *e.g.*, after the padding, the 384-fu supercell (21.4 x 21.4 x 33.1 Angstrom$^3$) now contains 6144 atoms and has a dimension of 42.7 x 42.7 x 66.3 Angstrom$^3$. This is to minimize the long-range periodicity error of the finite-size supercell. In Table 3 we list all generation parameters of the calculated PDF shown in Fig. 2 in the main text.

**The overall weighted R-value.** We use a weighted agreement factor $R_w$ to assess the agreement between calculated and observed PDF, which is given by

$$R_w = \sqrt{\frac{\sum_{i=1}^{n}[g_{obs}(r_i) - g_{calc}(r_i, P)]^2}{\sum_{i=1}^{n}[g_{obs}(r_i)]^2}}, \qquad (1)$$

**Appendix C: Effective band structure**

The basic concept of EBS can be described using the following equations. Assume in supercell $|Km\rangle$ is the m-th electronic eigen state at $K$ in supercell BZ whereas in primitive cell $|k_in\rangle$ is the n-th eigen state at $k_i$ in primitive BZ, then each $|Km\rangle$ can be expanded on a complete set of $|k_in\rangle$ where $K = k_i - G_i$, and $G_i$ being reciprocal lattice vectors in the supercell BZ, which is the folding mechanism [52]

$$|Km\rangle = \sum_{i=1}^{N_K} \sum_{n} F(k_i, n; K, m)|k_in\rangle, \qquad (2)$$

The supercell band structure at *K* can then be *unfolded* back to $k_i$ by calculating the spectral weight $P_{Km}(k_i)$

$$P_{Km}(k_i) = \sum_n |\langle Km|k_i n\rangle|^2 \tag{3}$$

$P_{Km}(k_i)$ represents 'how much' Bloch characteristics of wavevector $k_i$ has been preserved in $|Km\rangle$ when $E_n = E_m$. The EBS is then calculated by spectral function $A(k_i, E)$

$$A(k_i, E) = \sum_m P_{Km}(k_i)\delta(E_m - E) \tag{4}$$

**Appendix D: Peak analysis of α-γ bands in the calculated spectral function.**

The 3 bands α-γ near Fermi level are extracted from the calculated spectral function of the (Yes, Yes) PM supercell (2D colored contour in Fig. 6a, which is identical to the one shown in main text Fig. 1b), via the Lorentzian peak fit, as shown by the black dash lines in Fig. 6a,b, then compared to the ARPES peaks extracted from literature [34–36], shown as empty red circles in Fig. 1a,b and Fig. 6a.

**Acknowledgments**

We would like to thank Emil S. Božin for insightful discussions, and for sharing data from ref [41]. The work in the Zunger group and Billinge group was supported by the US National Science Foundation through grant DMREF-1921949. The calculations were done using the Extreme Science and Engineering Discovery Environment (XSEDE), which is supported by the National Science Foundation grant number ACI-1548562.

# Figures and Tables

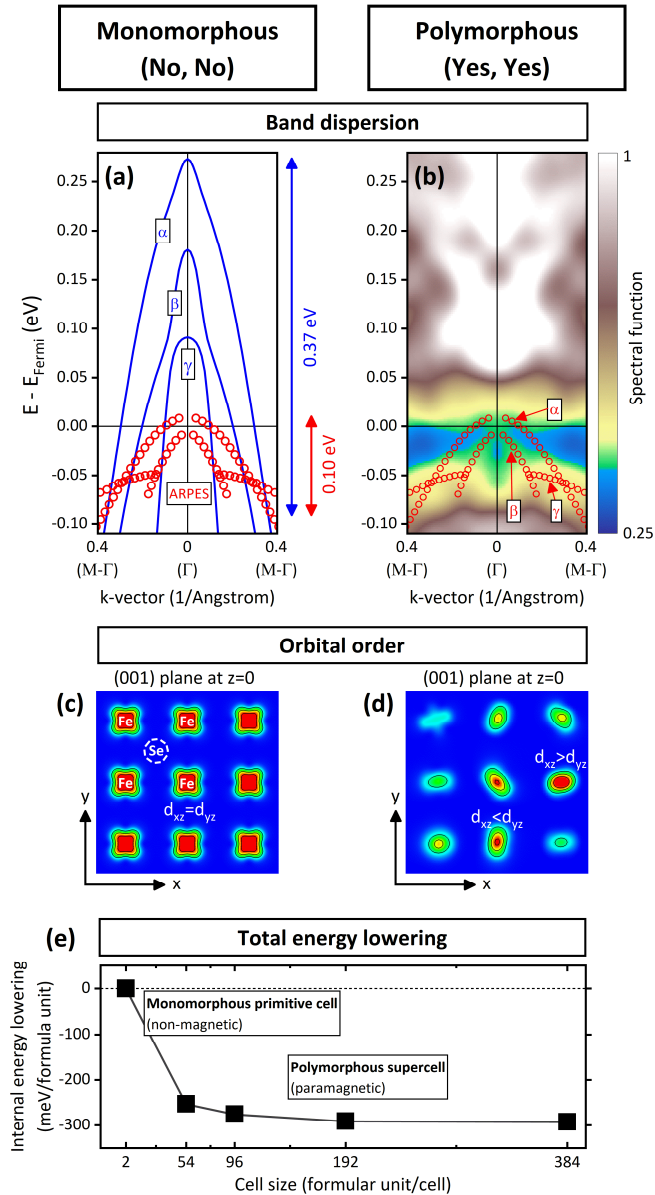

**Fig. 1 | Comparison of DFT-calculated properties of tetragonal FeSe** using the nominal monomorphous non-magnetic minimal-cell model (2 f.u./cell, part (a),(c)), having a single, repeated structural motif, and the polymorphous paramagnetic supercell model (384 f.u./cell, part (b),(d)), allowing a distribution of displacements and spin local environments. All calculations include spin-orbit coupling. The label in the title [(No, No) and (Yes, Yes)] denotes the existence or not of the spin and positional polymorphism, respectively. (a) and (b): Band structures compared with experimental ARPES results (red open circles; from Ref [34–36]) with α, β, and γ denoting three bands near the Fermi level. (c) and (d): Orbital order, drawn for bands within specific energy range of -0.03 eV < E - $E_F$ < 0.02 eV in (a) and (b), showing the *real-space* distribution of the cross-section

(here, (001) plane at z=0) of the partial charge density of such bands. Here (c) shows an identical four-fold symmetric charge distribution around every Fe sites (atoms positions are shown by circles and labels), implying equal occupation of $d_{xz}$ and $d_{yz}$ orbitals, while (d) shows an atomic-site-dependent, symmetry-broken charge distribution, indicating a local orbital ordering *i.e.* a local nematicity. (e): Lowering of total internal energy as cell size increasing. The size of Fermi surface pockets can be read from the EBS in (b), which is as small as the ARPES observation (open circles in (b)).

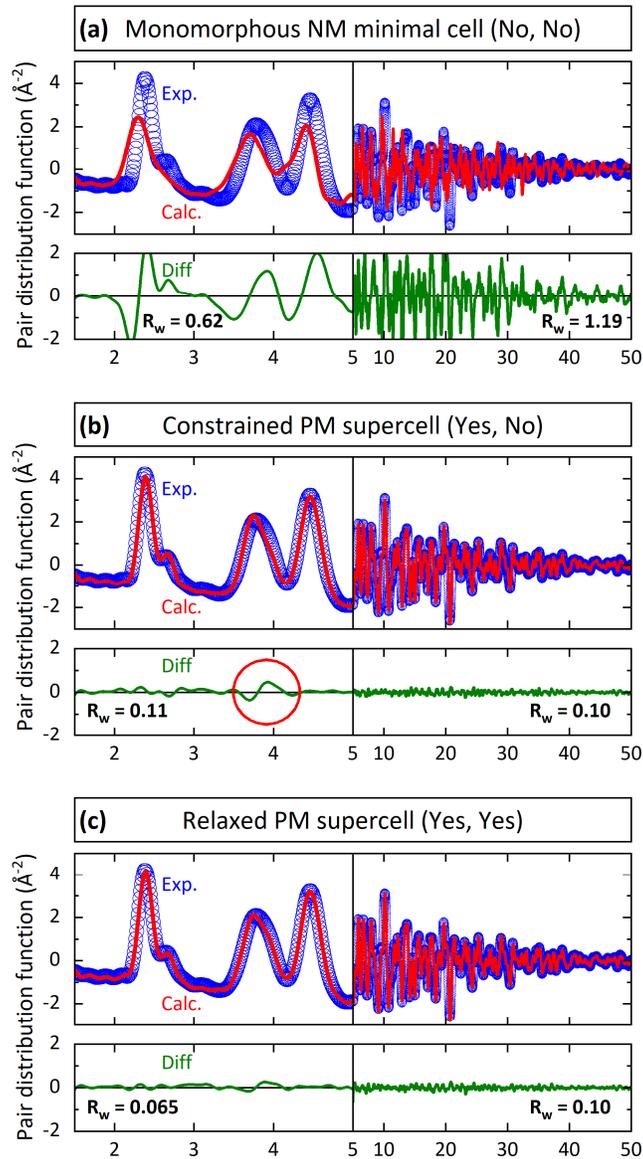

**Fig. 2 | Comparison between an experimentally measured PDF at room temperature** [41] **(blue empty circles) and DFT predicted PDF (red lines)**, shown together with the difference (green lines) for (a) monomorphous NM minimal-cell model (tetragonal, 2 f.u./cell), (b) constrained polymorphous PM supercell in the absent of atomic displacement (tetragonal, 384 f.u./cell), and (c) relaxed polymorphous PM supercell in the present of atomic displacement (tetragonal, 384 f.u./cell). The overall R-factors are also given. Red circle in (b) shows the disagreement around 4 Å.

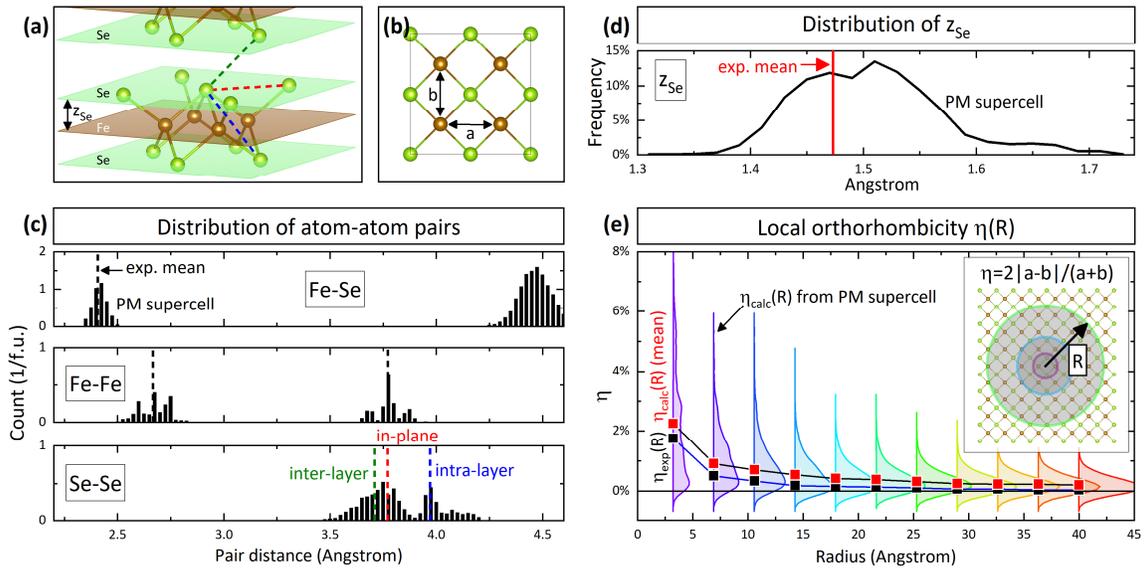

**Fig. 3 | Structural parameters characterizing the polymorphous network comprising the tetragonal (Yes, Yes) polymorphous PM supercell.** (a) Tetragonal FeSe has a layered structure and 3 unique Se-Se pair distance (in-plane, red dash line; intra-layer, blue; inter-layer, dark green). (b) The top view of monomorphous tetragonal FeSe shows two equivalent lattice constants a=b. (c) The calculated distribution of atom pairs in the PM supercell (black histograms), compared to the experimental observed mean values (vertical dash lines); notice the strong non-Gaussian distributions for Fe-Fe and Se-Se distances. (d) The calculated selenium height $z_{Se}$ in the PM supercell (black curve), compared to the experimental observed mean value (red vertical line). (e) The calculated local orthorhombicity $\eta_{calc}(R)$ (color-filed curves) from the supercell, and their local mean (red square) compared with the experimental local means (black squares) from Ref [41].

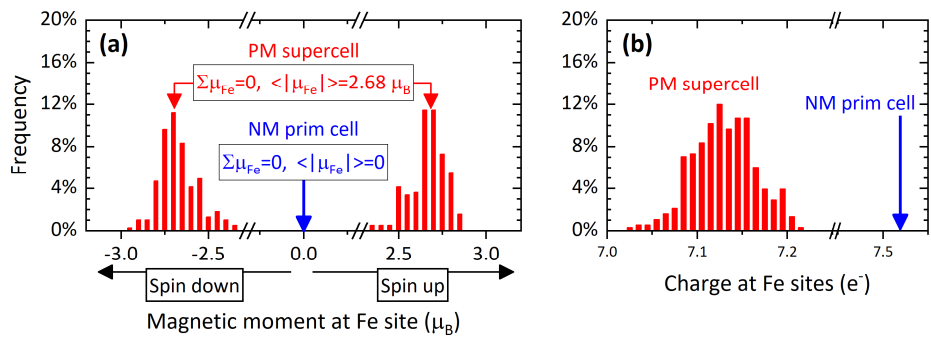

**Fig. 4 | Histogram plots for (a) Local magnetic moment and (b) local electronic charge at Fe sites** obtained from (Yes, Yes) PM supercell (red bars), compared with the single values from (No, No) NM minimal-cell model shown by blue lines.

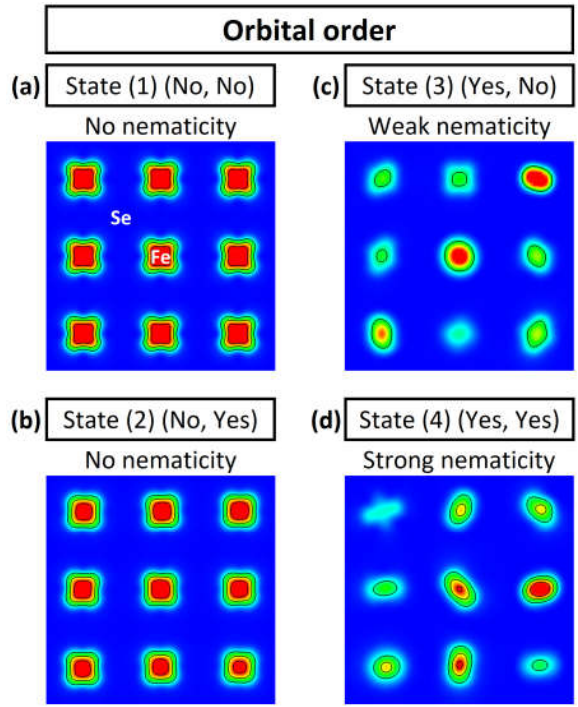

**Fig. 5 | The orbital orders for the 4 states in Table 1 and Table 2**, drawn for bands within specific energy range of -0.03 eV < E - EF < 0.02 eV in the electronic band structure, showing the real-space distribution of the cross-section (here, (001) plane at z=0) of the partial charge density of such bands. (a) the (No, No) state and (d) the (Yes, Yes) state are identical to the ones already shown in Fig. 1. The panels (a) and (b) show no significant symmetry breaking of the 4-fold rotational symmetry, indicating an equal occupation of $d_{xz}$ and $d_{yz}$ orbitals, hence have no nematicity; (c) shows a weak nematicity, where the partial charge density at different Fe sites have slightly different x and y distributions; (d) shows the strongest nematicity among the 4 states.

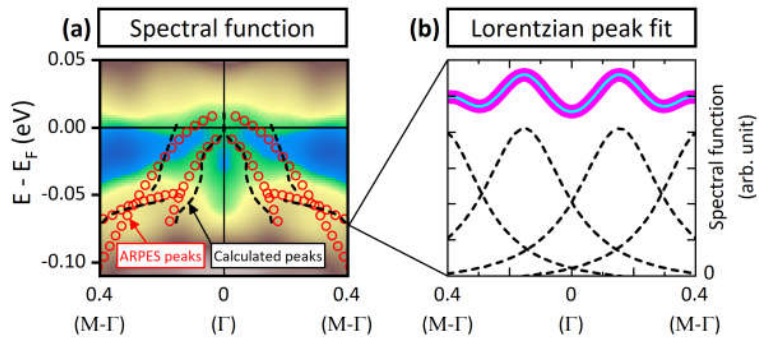

**Fig. 6 | The Lorentzian peak fit for the α-γ bands near Fermi level.** (a) The comparison between the calculated raw spectral function of the (Yes, Yes) PM supercell, which is identical to Fig. 1b, the ARPES measurement [34–36] (red circles), and the peaks extracted from the calculated spectral function (black dash lines). (b) One example of the peak analysis, for the calculated spectral function along M-Γ-M path at $E = E_F - 0.07$ eV in (a): The calculated raw spectral function is shown by the thick magenta line, while the 4 Lorentzian functions used to fit the spectral function are shown by black dash lines; cyan line is the cumulative result of the 4 Lorentzian peaks, which is in good agreement with the spectral function.

**Table 1 | Effects of the two polymorphous networks (spin and positional) in tetragonal FeSe**, on the total energy lowering ($\Delta E_{tot}$, state (1) has been chosen as the reference), the error to experimental PDF [41], the maximum of local orthorhombicity [41], and the existence or not of the electronic nematicity.

|  |  | Non-magnetic | | Para-magnetic | |
|---|---|---|---|---|---|
|  |  | (1) | (2) | (3) | (4) |
| **Nature of polymorphism** | Spin polymorphous | No | No | Yes | Yes |
|  | Positional polymorphous | No | Yes | No | Yes |
| **Consequence of polymorphism** | $\Delta E_{tot}$ (meV/fu) | 0 | -8 | -255 | -292 |
|  | PDF error ($R_w$) | 62% | 62% | 11% | 6.5% |
|  | Local orthorhombicity $\eta(R)$ | 0 | 0.92% | 0 | 2.2% |
|  | Electronic nematicity | No | No | Weak | Yes |

**Table 2 | Parameters and properties of different FeSe models.** Magnetic order: NM = nonmagnetic; PM = paramagnetic. Fully relaxed: all atomic positions as well as lattice constants (keeping tetragonal, *i.e.*, a=b≠c, α=β=γ=90°) are optimized by DFT total energy minimization. Relaxed internally: all atomic positions are optimized by DFT total energy minimization, while the lattice constants are fixed -- NM lattice constants are identical to the ones in state 1 (No, No), while PM lattice constants are taken from the full relaxation of a 8-f.u. PM supercell (which gives a bulk lattice constant of a=5.338 Å and c=5.521 Å conveniently close to the experimental value at room temperature a=5.334 Å and c=5.524 Å at 300 K [41]). Cell size is in number of formula units per cell. Crystallographic parameter $z_{Se}$ is the distance in [001] direction between Se layer and center Fe layer, given also as the Wyckoff position of Se atom in fractional coordinate. DFT total energies are relative to reference (state 1).

|  | Model | Cell size (f.u./cell) | *a, b, c* (Å) | $z_{Se}$ (Å), Wyckoff | DFT total energy (meV/f.u.) |
|---|---|---|---|---|---|
|  | Experimental tetragonal | - | 5.334, 5.334, 5.524 [a] | 1.473, 0.267 [a] | - |
| NM tetragonal | State 1 (No, No) (Fully relaxed) | 2 | 5.138, 5.138, 5.452 | 1.401, 0.257 | 0 (ref) |
| NM tetragonal | State 2 (No, Yes) (Relaxed internally) | 192 | 5.138, 5.138, 5.452 | 1.401, 0.257 | -8 |
| PM tetragonal | State 3 (Yes, No) (Not relaxed) | 384 | 5.338, 5.338, 5.521 | 1.47, 0.270 | -255 |
| PM tetragonal | State 4 (Yes, Yes) (Relaxed internally) | 384 | 5.338, 5.338, 5.521 | 1.48, 0.270 | -292 |

**Table 3 | Generation parameters of the calculated PDF shown in Fig. 2 in the main text:** (a) monomorphous NM minimal-cell model (No, No), (b) constrained PM supercell (Yes, No), and (c) relaxed PM supercell (Yes, Yes).

|  | Scaling factor | $\delta_1$ | ADP of Fe (U11, U22, U33) ($Å^{-2}$) | ADP of Se (U11, U22, U33) ($Å^{-2}$) |
|---|---|---|---|---|
| (a) Monomorphous NM | 0.755 | 1.481 | 0.032, 0.032, 0.0002 | 0.037, 0.037, 0.00002 |
| (b) Constrain PM | 0.800 | 1.896 | 0.015, 0.015, 0.024 | 0.014, 0.014, 0.011 |
| (c) Relaxed PM | 0.674 | 1.911 | 0.018, 0.018, 0.009 | 0.023, 0.023, 0.00001 |